# Khufu, Khafre and Menkaure Pyramids and the Sun

**Amelia Carolina Sparavigna**

Department of Applied Science and Technology, Politecnico di Torino, Torino, Italy

**Abstract**: In this paper we discuss the orientation of the Egyptian pyramids at Giza with respect to sunrises and sunsets, using SunCalc.net software. We can see that Khufu and Khafre pyramids had been positioned in a manner that, from each pyramid, it was always possible to observe the points of the horizon where the sun was rising and setting on each day of the year. A discussion for the Menkaure pyramid is also proposed.

The Great Pyramid, which is also known as the Pyramid of Khufu (Cheops) is the oldest and largest of the three pyramids at Giza, Egypt. It is one of the Seven Wonders of the Ancient World, the only one that survived till our days largely intact. The pyramid is the funeral monument of the fourth dynasty Egyptian Pharaoh Khufu, built as his tomb over a 10 to 20-year period concluding around 2560 BC [1]. Several scientific and alternative theories about the Great Pyramid's construction techniques had been proposed, with the most accepted construction hypotheses based on the idea that the pyramid was built by moving stones from a quarry and dragging and lifting them into place [2]. Inside the Great Pyramid, there are three known chambers, corridors, passages and shafts [3]. On the shafts as passages for the King's soul to the stars, it had been written in [4-6].
Other two pyramids are at Giza, the pyramids of Khafre and Menkaure. The Pyramid of Khafre (Chephren) is the second-tallest and second-largest of the pyramids of Giza, tomb of the pharaoh Khafre, who ruled from c. 2558 to 2532 BC, son of Khufu [7]. The smallest of the three pyramids is that of Menkaure, the last that had been built. It served as the tomb of Pharaoh Menkaure, son of Khafra and grandson of Khufu.
The remarkable monumental environment of these three pyramids led to several hypothesis on their planning. One is connecting the three pyramids to Orion stars [8]. In [9], the planning was also connected to an original project of the pyramid complex of Khufu at Giza as including also the project of the second pyramid complex, that of Khafre, at the same site. In paper [9], it is shown solar alignment at sunset in reference to the Great Sphinx of Giza, for the Khafre pyramid. About this alignment, a concern could arise: the Sphinx is facing East, whereas the alignments are toward sunsets. The same researcher had proposed alignments of pyramids to the site of Heliopolis [10].
For what concerns a solar alignment of the pyramids, let us consider an approach based on software which is giving the direction of sunrise and sunset on Google Earth satellite maps. Such software is provided by SunCalc.net and sollumis.com. They had been used for studying the orientation of the planning of several monumental sites [11-16]. The use of SucCalc was first proposed in [17,18].
Here we use SunCalc in the same manner we used sollumis.com for the Chinese Pyramids in [19]. At the Chinese burial places, we have usually two pyramids, one for the Emperor and the other for his Empress, which seem linked by the light of the sun at sunrise and sunset on solstices. Are the Giza pyramids linked by the sun, as the Chinese pyramids are? Let us investigate a possible answer. Before starting the discussion, let us remember that the pyramids of Giza have a square basis, with sides perfectly parallel to the cardinal direction North-South and East-West. The first pyramid was that of Khufu: probably a gnomon, with its shadow, was used to find the North-South cardinal direction. This direction was used to orient the pyramid, built in a place where the materials for building it was abundant, on the top of the Giza Plateau. The pyramid had a free horizon about it and its King, Khufu, could see the points of his local horizon where the sun was rising and setting on each day of the year.

The second pyramid that had been built was that of Khafre. In fact, Khafre had probably the same desire to admire sunrise and sunset from its pyramid. But its horizon was not free because constrained by his father's pyramid. It is probable that Khafre's architects decided the position of the new pyramid, in order that the king could see the points of the horizon where the sun was rising and setting thorough the year, maintaining the same possibility for Khufu. And in fact, if we use SucCalc (see Figure 1), we can see that Khufu sees the point of the horizon where the sun is setting on the Winter Solstice and therefore he can see the same on each days of the year. This happens because Khafre's architects have considered a proper place for the new pyramid. No problems for sunrise, because horizon is free. Of course, the Khafre's architects have considered that their king would like to observe the sunrise on each day of the year. Using SunCalc (see Figure 2) we can find that Khafre can see the point of the horizon where the sun is rising on the summer solstice, and then he can see the same on each day of the year. No problems for sunset, because horizon was free.

After, a third pyramid was added. This is the Menkaure's pyramid. It is smaller. If we consider the point at the North-West corner of this Pyramid, we have that, from it, we can see the point of the horizon where the sun is rising on summer solstice. If we imagine a larger pyramid (red square), and the corner as it center, we could repeat the same discussion we made for the pyramids of Khufu and Khafre, in the Figures 2 and 3.

In this paper, we have proposed a simple approach to the planning to the Giza pyramids. Using SunCalc we observe that, from each pyramid, the king's soul, imagined at the center of his pyramid, could see the points of the horizons where the sun was rising and setting thorough the year.

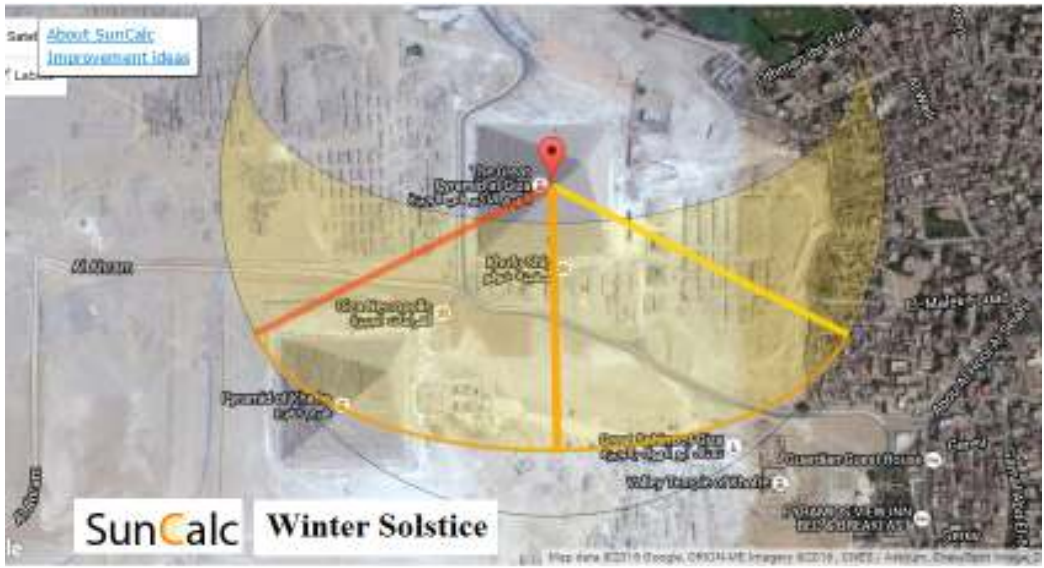

Figure 1: Using SunCalc we can find that Khufu can see the point of the horizon where the sun is setting on the winter solstice, and then he can see the same on all the days of the year. This happens because Khafre's architects have considered a proper place for the new pyramid. No problems for sunrise, because horizon is free.

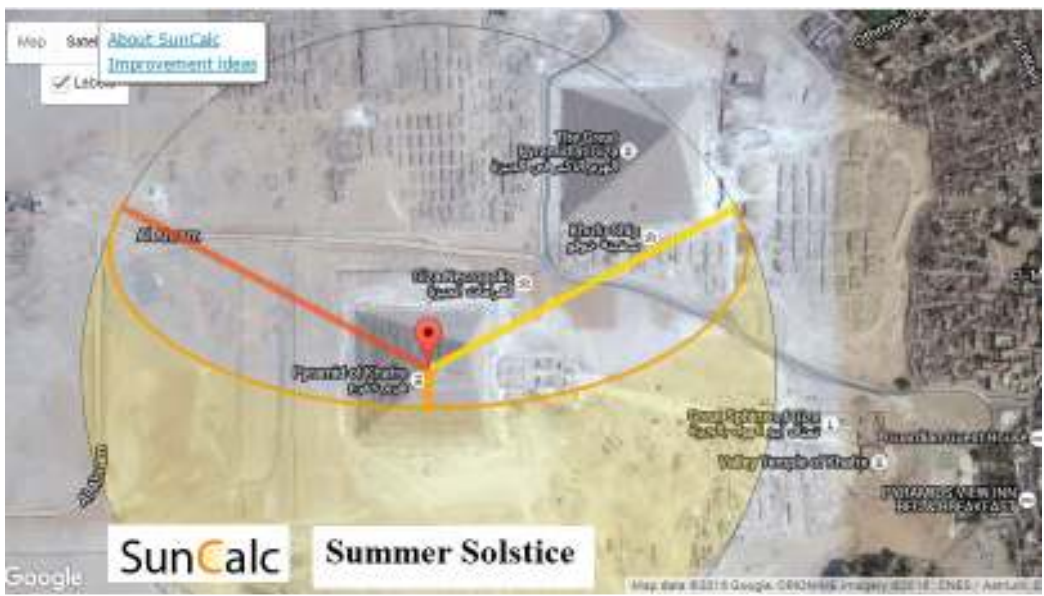

Figure 2: Of course, Khafre's architects have considered that their king would like to observe the sunrise on each day of the year. Using SunCalc we can find that Khafre can see the point of the horizon where the sun is rising on the summer solstice, and then he can see the same on all the days of the year. No problems for sunset, because horizon was free.

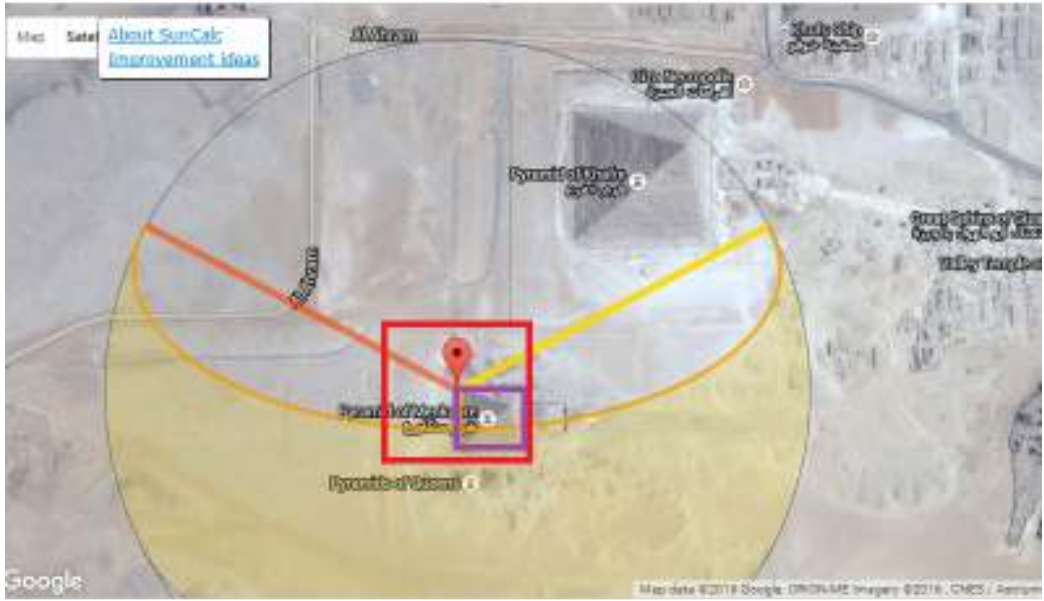

Figure 3: The third pyramid is that of Menkaure. It is smaller. If we consider the point at the North-West corner of this pyramid, we have that, from it, we can see the point of the horizon where the sun is rising on the Summer Solstice. If we imagine a larger pyramid (red square), and this corner as it center, we could repeat the same discussion as given in the Figures 2 and 3.